Financial and symbolic incentives promote 'green' charging choices

Kacperski, Celina[1&2] & Kutzner, Florian[1&2]

University of Mannheim[1] , University of Heidelberg[2]

[1]University of Mannheim, Department of consumer and economic psychology, Parkring 47, Raum 313, 68159 Mannheim, Germany. [2]University of Heidelberg, Department of cognitive research in social psychology, Hauptstr. 47-51, 69117, Heidelberg, Germany.

e-mail: kacperski@uni-mannheim.de

Acknowledgements

Financial support by the European Union Horizon 2020 research and innovation programme is gratefully acknowledged (Project ELECTRIFIC, grant N°713864).\

February 2020

Please find published version here:

Abstract

Electromobility can contribute to a reduction in greenhouse gas emissions if usage behavior is aligned with the increasing availability of renewable energy. To achieve this, smart navigation systems can be used to inform drivers of optimal charging times and locations. Yet, required flexibility may impart time penalties. We investigate the impact of financial and symbolic incentive schemes to counteract these additional costs. In a laboratory experiment with real-life time costs, we find that monetary and symbolic incentives are both effective in changing behavior towards 'greener' charging choices, while we find no significant statistical difference between them.

Transportation accounts for 28% of $CO_2$ emissions in Europe, in large part due to private vehicle use (Capros & Europäische Kommission, 2014). With electric vehicles (EVs) on the rise (Vaughan, 2017), it is time to scrutinize how the usage behavior of EVs can contribute to reducing $CO_2$ emissions (Buekers, Van Holderbeke, Bierkens, & Int Panis, 2014). The largest share of an EV's life time emissions is caused by the production of the charged electricity (Frischknecht, Messmer, & Stolz, 2018). Thus, aligning charging behavior with renewable electricity production is key to reduce the direct carbon footprint of EVs (Richardson, 2013; Schmalfuß et al., 2015; Sundstrom & Binding, 2012). Additionally, successful alignment will reduce EVs' indirect ecological footprint by shaving peak demand and thus reducing the necessity for grid extensions (Green, Wang, & Alam, 2011).

Previously, research has focused on financial incentives for steering charging behavior (Dallinger & Wietschel, 2012; Flath, Ilg, Gottwalt, Schmeck, & Weinhardt, 2013; Li, Wu, & Oren, 2014). Yet little is known about the behavioral demand elasticity for EV charging prices. Existing evidence for home electricity consumption points towards variable and low elasticities (Azevedo, Morgan, & Lave, 2011). For the EU, elasticities are around -.20, i.e. a 20% increase in price will reduce consumption by about 1%. This is in line with findings on purely financial incentives in other domains, where for example low and variable effects have been documented for performance in laboratory tasks (Bonner, Hastie, Sprinkle, & Young, 2000), for fitness and physical activity (Barte & Wendel-Vos, 2017) and for unhealthy food consumption (Cornelsen et al., 2015). To the best of our knowledge, ours is the first study investigating the impact of changes in incentives on decision making and behavior in the context of EV charging.

Further, especially for EV charging, non-financial or symbolic incentives should be considered as an option. EV owners have been found to be strongly motivated by ecological factors (Gaker, Vautin, Vij, & Walker, 2011; White & Sintov, 2017), and environmental performance as well as symbolic attributes surpass instrumental aspects such as price and range as predictor of EV purchase intentions (Degirmenci & Breitner, 2017). Symbolic incentives for environmentally friendly behaviours have been investigated in the form of gamification, fostering consumption awareness and increasing energy-savings (Morganti et al., 2017), informational or normative messages, which decreased energy usage by up to 6%, and reduced littering (Allcott, 2011; Cialdini, Reno, & Kallgren, 1990) and emotional incentives, for example a polar bear animation on a melting ice sheet, which reduced the consumption of hot water while showering (Tiefenbeck, Tasic, Schob, & Staake, 2013). Yet, in few of these studies, choices have obvious time costs for participants, and none were in the context of charging EVs.

Further, the effects of financial incentives on behaviors have rarely been compared to those of symbolic incentives. Where they have been, results were mixed: the impact of meaning and recognition was found to be stronger on effort and performance in a data-entry job context than financial rewards (Kosfeld, Neckermann, & Yang, 2017); in a large educational field experiment, both outright monetary reward and a travel opportunity of similar value increased exam scores equally (Burgess, Metcalfe, & Sadoff, 2016).

In the context of energy savings and environmental benefits, one study on household electricity consumption found financial incentives to be stronger than comparative social feedback (Mizobuchi & Takeuchi, 2013), while another found the opposite, namely that emphasizing monetary as compared to environmental benefits of enrolling in an energy-savings program reduced their willingness to enroll (Schwartz, Bruine de Bruin, Fischhoff, & Lave, 2015). Similarly, intentions to adopt energy-saving driving behaviors (e.g. turning off the engine while idling, checking tire pressure) were less responsive to information on the financial savings they entail as compared to information about environmental benefits (Dogan, Bolderdijk, & Steg, 2014). Compliance to an economic savings appeal as opposed to an environmental appeal was found to be lower in an online study that introduced a tire pressure check scenario at a gas station (Bolderdijk, Gorsira, Keizer, & Steg, 2013). We firstly explore this in the context of EV charging. Commercial actors such as BMW, fleetcarma, and various utility providers including EDF, E.ON and Enel are showing strong interest in incentivizing charging with both financial and symbolic incentives, ("BMW ChargeForward," 2020; "Charge the North," 2019; "Electric vehicle infrastructure," 2019; Whited, Avi, & Wilson, 2018), so scientific evidence comparing the effectiveness of monetary and symbolic incentives for flexible charging of EVs will be paramount to implementation of effective programs.

A pragmatically feasible way of delivering incentives in the context of EV driving and charging is via a smartphone-based navigation application. We used an app (see Figure 1) that was designed to display information about ecologically friendly driving and charging alongside routing information (Electrific, Eider et al., 2017; Kacperski & Kutzner, 2017). By adding information about incentives, we investigated the flexibility of charging choices.

More specifically, we investigated the number of "green" charging decisions made in a laboratory setting involving a routing task. The task offered the choice to select a charging station due to currently available green energy, with a time penalty for this "green" choice. The time penalty represents time lost due to detours for navigating to additional charging points.

We predicted a main effect of both financial and symbolic incentives (as compared to no incentive), expecting that both incentives would significantly increase the number of "green" choices. Additionally, we explored the effect of various moderators such as gender, environmental attitudes and instrumental and symbolic attitudes towards EVs.

## Methodology

### Participants

Our convenience sample comprised 173 students (121 women, 15 other), recruited to the laboratory in mid-2018. Mean age was 24.52 ($SD$ = 7.78) of which 32.5% reported EV experience. Participants were recruited via a university online study portal and informed that they could receive up to 7 Euro variable compensation, depending on the choices they made during the session. When participants arrived at the laboratory, they signed a letter of information and consent form; then they were seated at one of the computers. First, participants filled out a survey, then received the smartphone equipped with the ELECTRIFIC navigation application (see Figure 1), and a sheet with instructions on their required navigation for the study. Once finished, they returned to the PC to complete a cognitive attention task to simulate the time spent driving given their selected green or fast route.

### Materials

#### Questionnaire.

The questionnaire consisted of items generated to gain information about a participants' EV experience as well as environmental attitudes and attitude functions (for all items, see Appendix). Participants were asked via five items: (1) whether they owned an EV and/or (2-5) had used one in the context of car rentals, car sharing, by using a shared neighbor vehicle, or as a family vehicle (each answer coded 0,1). A positive answer in any one of these items resulted in the judgement that the participant had EV experience (coded 1,0). One item measured EV purchase intentions, another two measured EV attractiveness on a 5-point Likert scale (disagree to agree). Environmental attitude was measured via a selection of the four best loading items

from the New Environmental Paradigm (Stern, Dietz, & Guagnano, 1995) on a 5-point Likert

scale (disagree to agree). Mean environmental attitude in our sample was high at 4.4 ($SD = 0.6$, α

= .74).

We surveyed symbolic and instrumental attitude functions related to EV usage (Shavitt,

1990) on a 5-point Likert scale (disagree to agree). Symbolic items included  "I would drive an

electric car, because I can showcase my values with this; … because I can be part of the

sustainability movement" and instrumental items it is easier to drive due to its quick acceleration

and its silent motor; … because it can save me time due to home/destination charging". The

symbolic and instrumental attitude scales had satisfactory reliabilities of α = .74 ($M = 3.6$, $SD =$

0.65), and α = .73 ($M = 2.3$, $SD = 0.77$), respectively.

**ELECTRIFIC smartphone application**

Once participants completed the survey section, they received a smartphone (LG K8)

with the opened ELECTRIFIC application; this app allows EV drivers to plan and navigate a trip, along with charging station suggestions for steered "green" charging (see Figure 1).

Participants received an instruction manual that gave them a start-destination pair and a cover story. They were to imagine they were driving their EV from home to the train station in the city, a trip of about 4km, and could choose the fastest routing option by parking there directly without navigating to a nearby charging station, or the green routing option – due to currently available renewable energy – , including the required navigation to the nearby charging station, which imparted a time penalty[1]. They were informed that once they started navigation, they would be moved to the PC to spend the complete time of the trip on a task simulating the driving of the route. After entering the start and destination locations, participants saw the selection screen (see Figure 1) and were assigned to one of the three conditions.

_______________________

[1] Note that even if participants selected the fast routing option, they could still actively include
the charging station in the route. None of the participants chose to do so.

In the control condition, only the choice dropdown without a preselected option and without additional information was present, allowing participants to select the "green" and the "fast" route. The design choice for the "green" route clearly marked it as more eco-friendly (see Figure 2). In the symbolic incentive condition, participants saw a figure with information about the $CO_2$ emission difference between green and fast route, along with information that made the difference evaluable (Hsee & Zhang, 2010), transforming the amount of $CO_2$ in the equivalent of cups of coffee (Zapico, Guath, & Turpeinen, 2011). A note at the bottom of the screen informed them that the reduced emission cost was due to steered charging. After 300ms, a pop-up notification informed participant that all $CO_2$ "saved" in the study would be compensated via the non-profit $CO_2$ offset organization Atmosfair. In the financial incentive condition, participants were shown the cost difference between green and fast route, again with the note that reduced cost was due to the steered charging. A pop-up informed participants that the increase in cost of the fast route (random assignment of 1€, 1.50€ and 2€) would be deduced from their participation compensation at the end of the study. Therefore, participants in the control and in the symbolic group received 7 Euro compensation at the end of the trip; participants in the financial incentive condition received between 5 and 7 Euro depending on their selection.

Participants, once they selected one of the routes, were forwarded to a map of the trip, including a route summary that showed them the km and time of the trip in real time (see Figure 1b). Resulting average durations where 7.0 min ($SD = 0.17$) for the fast and 13.3 min ($SD = 1.3$) for the green route. They had the opportunity to return to the previous screen and reselect the other route if they wished. Once they had decided, they started "navigation" via a button and handed the smartphone to the research assistant and returned to the PC. They had been informed that as soon as their "driving time" was up, they could collect compensation and leave the laboratory.

**"Driving" task.**

We used the Attention Network Test (CRSD-ANT, Fan, McCandliss, Sommer, Raz, &

Posner, 2002) , a tool validated for driving score predictions as it targets functions of attention relevant in driving situations (Weaver, Bédard, McAuliffe, & Parkkari, 2009). The purpose of this task was to create an appropriate cost for the selection of a green versus fast route that is similar to the psychological cost while driving; participants selecting the fast route had to spend approximately 45% less time in the laboratory on a tedious task that still required their full attention, as for each error, they lost 5 cents of their compensation money.

**Intervention and dependent variable**

For our one-factorial design with three levels, our main analysis comprised a comparison of choices made in conditions of no incentive to symbolic or financial incentives. Each participant was assigned to one of the conditions at random at the beginning of the study. For exploratory purposes, we included a number of sublevels to the symbolic and financial incentives by varying their strength. In the financial condition, we varied participants' cost between 1€,

1.50€ and 2€. In the symbolic condition, we varied the amount of $CO_2$ as 1.7kg, 2.2kg and 2.7kg,

represented by 8, 10 and 12 cups of coffee. As a dependent variable, we measured the binary choice of the green or the fast route.

 Analyses were computed a generalized linear model using a Helmert and the glm command, binomial family, of the stats package in R (R Development Core Team, 2008).

## Results

For our main analyses, we compared the control condition, coded -2, to the monetary and symbolic incentive conditions, each coded +1. Orthogonally, we also compared the monetary incentive condition, coded – 1, to the symbolic incentive condition, coded +1. We included predictors for gender, environmental attitudes (centered) and instrumental and symbolic attitudes towards EVs (centered), EV experience, as well as their interactions with the Helmert contrast.

The analysis revealed a significant difference between the control and incentive conditions, $z = 2.16$, $p = .03$, $OR = 3.24$ (CI: 1.29, 12.14), but no evidence for a difference between the two incentive conditions, $z = 0.98$, $p = .15$, $OR = 0.37$ (CI: 0.07, 1.30), with $M_{control} = 0.63$ ($SD = 0.48$) vs $M_{symbolic} = 0.87$ ($SD = 0.33$) vs $M_{financial} = 0.90$ ($SD = 0.30$).

Being a woman as compared to a man increased the chance of selecting the greener option, $z = 2.69$, $p = .007$, $OR = 8.09$ (CI: 1.80, 43.12), with $M_w = 0.89$ ($SD = 0.32$) vs $M_m = 0.69$ ($SD = 0.47$). None of the other effects were significant $|z| < 1.89$, $p > .06$.

We also explored whether various magnitudes of the incentives would differ in their effect: we predicted choices from the magnitude values, the type of incentive, financial or symbolic, and their interaction. Contrasts were coded as -1, 0, 1 for magnitude, and -1,1 for financial vs symbolic. As before, we included predictors for gender, environmental beliefs, and instrumental and symbolic attitudes towards EVs, as well as EV experience. For this analysis, the control group without incentives was excluded. None of the effects were significant, with $|z| < 0.56$, $p > .57$, with $M_{symbolic\_1} = 0.96$ ($SD = 0.21$), $M_{symbolic\_2} = 0.80$ ($SD = 0.41$), $M_{symbolic\_3} = 0.85$ ($SD = 0.37$) vs $M_{financial\_1} = 0.88$ ($SD = 0.32$), $M_{financial\_2} = 0.86$ ($SD = 0.35$), $M_{financial\_3} = 0.96$ ($SD = 0.20$).

## Discussion

We investigated whether symbolic incentives ($CO_2$ compensation with the amount made evaluable via coffee cups) or financial incentives (provided via participant compensation docking) would affect participant's selection of green vs fast route as compared to a control condition. We found participants in symbolic and financial incentive conditions chose the green route over 20% more often, despite higher time costs due to the longer "drive and charge" associated with it.

We found that symbolic and financial incentives both worked, but we found no significant statistical difference between them. One explanation could be that from the perspective of participants, the cost of $CO_2$ versus Euro holds similar persuasive weight and therefore showed the same increase in green route selection. A second possibility, symbolic and financial incentives may have been perceived by participants as similar in their basic functionality, namely as information transmission strategies. Pricing as an incentive tool might here be a symbolic incentive in disguise, where the monetary cost is seen simply as a conversational vehicle to transmit relevant information – in the present case, cost in Euro communicating to individuals the added cost to the environment, therefore in meaning being equivalent to the symbolic incentive we represented via $CO_2$. This would be in line with the idea that individuals strongly depend on the communicative context to interpret information, in the manner of conversational implicature (Grice, 1975). This line of argument also finds support in the lack of effect of different levels of symbolic and financial incentives: small differences in costs (either $CO_2$ or Euro) did not affect the choice behavior in our sample; it will be worthwhile to investigate whether this is an indicator for a communication-level functioning of the two incentives rather than a weight-based one; here, the magnitude of the difference should also be further varied and increased.

Our findings have implications in the EV context, considering the strong focus on pricing strategies as a major driver to counteract uncoordinated charging (see Dallinger & Wietschel, 2012; Li et al., 2014). Certainly, beside economic factors, psychological parameters should be considered in user-inclusive modelling research of grid balancing and power distributions. Price elasticity might have to be estimated differently than thus far.

Some limitations deserve mentioning. Firstly, the participant number for analyses of varying incentive magnitudes was low, insufficient power possibly impeding the probability of finding significant effects between them. Secondly, the study was conducted in a laboratory setting with student participants, only a third of which were current EV users or had EV experience. However, as EV adoption rates are increasing, it is likely that many of the students queried here will become drivers of EVs in the near future. Therefore, gaining a first understanding of their decision-making in the context of electromobility is of some importance. Finally, under real circumstances, both benefits and costs of greener steered driving and charging might be much larger than those we chose in the present study; financially, choosing to charge when renewables are high might result in completely free charging, but might imply the higher cost of planning one's day around the availability of grid and power. Analyzing the trade-offs between financial and other costs in the face of financial and symbolic incentives is therefore a worthy subject of further study.

**References**

Allcott, H. (2011). Social norms and energy conservation. *Journal of Public Economics*, *95*(9),
        1082–1095. doi: 10.1016/j.jpubeco.2011.03.003

Azevedo, I. M. L., Morgan, M. G., & Lave, L. (2011). Residential and Regional Electricity
        Consumption in the U.S. and EU: How Much Will Higher Prices Reduce CO2
        Emissions? *The Electricity Journal*, *24*(1), 21–29. doi: 10.1016/j.tej.2010.12.004

Barte, J. C. M., & Wendel-Vos, G. C. W. (2017). A systematic review of financial incentives for
        physical activity: The effects on physical activity and related outcomes. *Behavioral
        Medicine*, *43*(2), 79–90. doi: 10.1080/08964289.2015.1074880

BMW ChargeForward. (2020). Retrieved January 3, 2020, from
        https://www.bmwchargeforward.com/#/home

Bolderdijk, J. W., Gorsira, M., Keizer, K., & Steg, L. (2013). Values Determine the
        (In)Effectiveness of Informational Interventions in Promoting Pro-Environmental
        Behavior. *PLoS ONE*, *8*(12), e83911. doi: 10.1371/journal.pone.0083911

Bonner, S. E., Hastie, R., Sprinkle, G. B., & Young, S. M. (2000). A Review of the Effects of
        Financial Incentives on Performance in Laboratory Tasks: Implications for Management
        Accounting. *Journal of Management Accounting Research*, *12*(1), 19–64. doi:
        10.2308/jmar.2000.12.1.19

Buekers, J., Van Holderbeke, M., Bierkens, J., & Int Panis, L. (2014). Health and environmental
        benefits related to electric vehicle introduction in EU countries. *Transportation Research
        Part D: Transport and Environment*, *33*, 26–38. doi: 10.1016/j.trd.2014.09.002

Burgess, S. M., Metcalfe, R., & Sadoff, S. (2016). *Understanding the Response to Financial and
        Non-Financial Incentives in Education: Field Experimental Evidence Using High-Stakes*

*Assessments* (SSRN Scholarly Paper No. ID 2861069). Retrieved from Social Science

Research Network website: https://papers.ssrn.com/abstract=2861069

Capros, P., & Europäische Kommission. (2014). *EU energy, transport and GHG emissions:*

*Trends to 2050 reference scenario 2013*. Luxembourg: Publications Office of the EU.

Charge the North: Results from the world's largest electric vehicle charging study. (2019).

Retrieved January 3, 2020, from FleetCarma website:

https://www.fleetcarma.com/resources/charge-the-north-summary-report/

Cialdini, R. B., Reno, R. R., & Kallgren, C. A. (1990). A focus theory of normative conduct:

Recycling the concept of norms to reduce littering in public places. *Journal of*

*Personality and Social Psychology*, *58*(6), 1015–1026. doi: 10.1037/0022-

3514.58.6.1015

Cornelsen, L., Green, R., Turner, R., Dangour, A. D., Shankar, B., Mazzocchi, M., & Smith, R.

D. (2015). What happens to patterns of food consumption when food prices change?

Evidence from a systematic review and meta-analysis of food price elasticities globally.

*Health Economics*, *24*(12), 1548–1559. doi: 10.1002/hec.3107

Dallinger, D., & Wietschel, M. (2012). Grid integration of intermittent renewable energy sources

using price-responsive plug-in electric vehicles. *Renewable and Sustainable Energy*

*Reviews*, *16*(5), 3370–3382. doi: 10.1016/j.rser.2012.02.019

Degirmenci, K., & Breitner, M. H. (2017). Consumer purchase intentions for electric vehicles: Is

green more important than price and range? *Transportation Research Part D: Transport*

*and Environment*, *51*, 250–260. doi: 10.1016/j.trd.2017.01.001

Dogan, E., Bolderdijk, J. W., & Steg, L. (2014). Making Small Numbers Count: Environmental

    and Financial Feedback in Promoting Eco-driving Behaviours. *Journal of Consumer*

    *Policy*, *37*(3), 413–422. doi: 10.1007/s10603-014-9259-z

Eider, M., Stolba, M., Sellner, D., Berl, A., Basmadjian, R., de Meer, H., … Kacperski, C.

    (2017). Seamless Electromobility. *Proceedings of the Eighth International Conference on*

    *Future Energy Systems  - e-Energy '17*, 316–321. doi: 10.1145/3077839.3078461

Electric vehicle infrastructure: Energy firms are taking an interest in EVs. (2019, October 25).

    Retrieved January 3, 2020, from Power Technology | Energy News and Market Analysis

    website: https://www.power-technology.com/comment/energy-suppliers-are-asking-what-

    can-ev-schemes-do-for-us/

Fan, J., McCandliss, B. D., Sommer, T., Raz, A., & Posner, M. I. (2002). Testing the Efficiency

    and Independence of Attentional Networks. *Journal of Cognitive Neuroscience*, *14*(3),

    340–347. doi: 10.1162/089892902317361886

Flath, C. M., Ilg, J. P., Gottwalt, S., Schmeck, H., & Weinhardt, C. (2013). Improving Electric

    Vehicle Charging Coordination Through Area Pricing. *Transportation Science*, *48*(4),

    619–634. doi: 10.1287/trsc.2013.0467

Frischknecht, R., Messmer, A., & Stolz, P. (2018). *Aktualisierung Umweltaspekte von*

    *Elektroautos*. 107.

Gaker, D., Vautin, D., Vij, A., & Walker, J. L. (2011). The power and value of green in promoting

    sustainable transport behavior. *Environmental Research Letters*, *6*(3), 034010. doi:

    10.1088/1748-9326/6/3/034010

Green, R. C., Wang, L., & Alam, M. (2011). The impact of plug-in hybrid electric vehicles on distribution networks: A review and outlook. *Renewable and Sustainable Energy Reviews*, *15*(1), 544–553. doi: 10.1016/j.rser.2010.08.015

Grice, H. P. (1975). Logic and conversation. In *Syntax and Semantics: Speech Acts* (pp. 41–58). Academic Press.

Hsee, C. K., & Zhang, J. (2010). General Evaluability Theory. *Perspectives on Psychological Science*, *5*(4), 343–355. doi: 10.1177/1745691610374586

Kacperski, C., & Kutzner, F. (2017). Assisting Ev Drivers To Act Smarter: How Behavioral Tools Can Facilitate Adoption Of Smart Navigation And Greenify Mobility. *Proceedings of E-Mobility Power System Integration Symposium*, *1*, 1–4. doi: 10.5281/zenodo.1161394

Kosfeld, M., Neckermann, S., & Yang, X. (2017). The effects of financial and recognition incentives across work contexts: The role of meaning. *Economic Inquiry*, *55*(1), 237–247. doi: 10.1111/ecin.12350

Li, R., Wu, Q., & Oren, S. S. (2014). Distribution Locational Marginal Pricing for Optimal Electric Vehicle Charging Management. *IEEE Transactions on Power Systems*, *29*(1), 203–211. doi: 10.1109/TPWRS.2013.2278952

Mizobuchi, K., & Takeuchi, K. (2013). The influences of financial and non-financial factors on energy-saving behaviour: A field experiment in Japan. *Energy Policy*, *63*, 775–787. doi: 10.1016/j.enpol.2013.08.064

Morganti, L., Pallavicini, F., Cadel, E., Candelieri, A., Archetti, F., & Mantovani, F. (2017). Gaming for Earth: Serious games and gamification to engage consumers in pro-environmental behaviours for energy efficiency. *Energy Research & Social Science*, *29*, 95–102. doi: 10.1016/j.erss.2017.05.001

R Development Core Team. (2008). *R: A language and environment for statistical computing.*

    Vienna, Austria: R Foundation for Statistical Computing.

Richardson, D. B. (2013). Electric vehicles and the electric grid: A review of modeling

    approaches, Impacts, and renewable energy integration. *Renewable and Sustainable*

    *Energy Reviews*, *19*, 247–254. doi: 10.1016/j.rser.2012.11.042

Schmalfuß, F., Mair, C., Döbelt, S., Kämpfe, B., Wüstemann, R., Krems, J. F., & Keinath, A.

    (2015). User responses to a smart charging system in Germany: Battery electric vehicle

    driver motivation, attitudes and acceptance. *Energy Research & Social Science*, *9*, 60–71.

    doi: 10.1016/j.erss.2015.08.019

Schwartz, D., Bruine de Bruin, W., Fischhoff, B., & Lave, L. (2015). Advertising energy saving

    programs: The potential environmental cost of emphasizing monetary savings. *Journal of*

    *Experimental Psychology: Applied*, *21*(2), 158–166. doi: 10.1037/xap0000042

Shavitt, S. (1990). The role of attitude objects in attitude functions. *Journal of Experimental*

    *Social Psychology*, *26*(2), 124–148. doi: 10.1016/0022-1031(90)90072-T

Stern, P. C., Dietz, T., & Guagnano, G. A. (1995). The New Ecological Paradigm in Social-

    Psychological Context. *Environment and Behavior*, *27*(6), 723–743. doi:

    10.1177/0013916595276001

Sundstrom, O., & Binding, C. (2012). Flexible Charging Optimization for Electric Vehicles

    Considering Distribution Grid Constraints. *IEEE Transactions on Smart Grid*, *3*(1), 26–

    37. doi: 10.1109/TSG.2011.2168431

Tiefenbeck, V., Tasic, V., Schob, S., & Staake, T. (2013). Mechatronics to drive environmental

    sustainability: Measuring, visualizing and transforming consumer patterns on a large

scale. *IECON 2013 - 39th Annual Conference of the IEEE Industrial Electronics Society*,

4768–4773. doi: 10.1109/IECON.2013.6699906

Vaughan, A. (2017, December 25). Electric and plug-in hybrid cars whiz past 3m mark

worldwide. *The Guardian*. Retrieved from

https://www.theguardian.com/environment/2017/dec/25/electric-and-plug-in-hybrid-cars-

3m-worldwide

Weaver, B., Bédard, M., McAuliffe, J., & Parkkari, M. (2009). Using the Attention Network Test

to predict driving test scores. *Accident Analysis & Prevention*, *41*(1), 76–83. doi:

10.1016/j.aap.2008.09.006

White, L. V., & Sintov, N. D. (2017). You are what you drive: Environmentalist and social

innovator symbolism drives electric vehicle adoption intentions. *Transportation Research

Part A: Policy and Practice*, *99*, 94–113. doi: 10.1016/j.tra.2017.03.008

Whited, M., Avi, A., & Wilson, R. (2018). *Driving Transportation Electrification Forward in

New York* (pp. 1–52). Synapse Energy.

Zapico, J. L., Guath, M., & Turpeinen, M. (2011). Kilograms or cups of tea: Comparing

footprints for better CO2 understanding. *PsychNology Journal*, *9*(1), 43–54.

# Figures

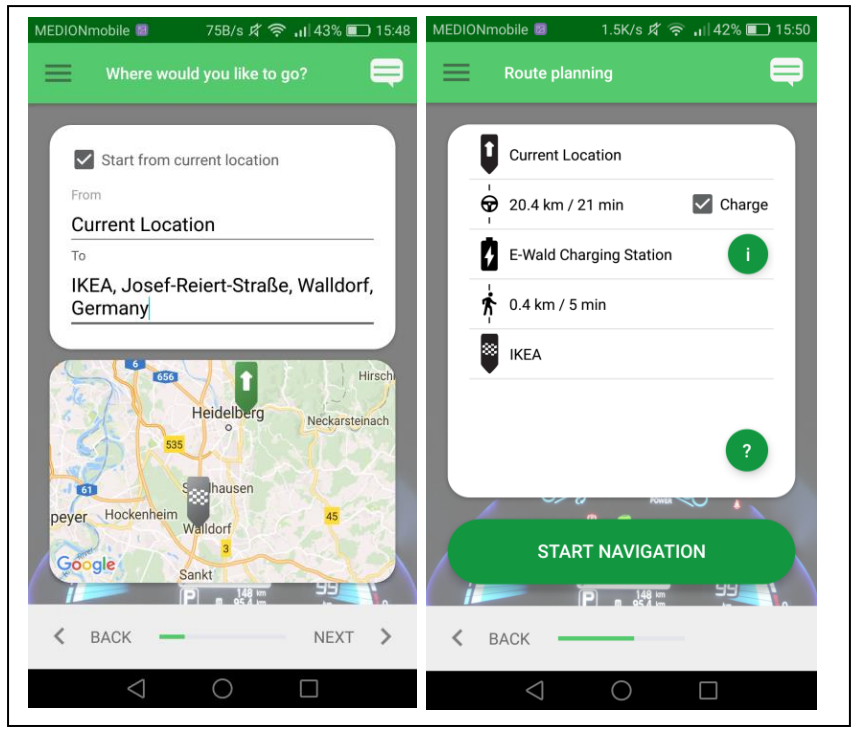

Figure 1. Electrific application. Left (1a): Example of a possible start-to-destination input text with map showing the trip. Right (1b): Route summary of a trip if a charging station is selected – "green" route.

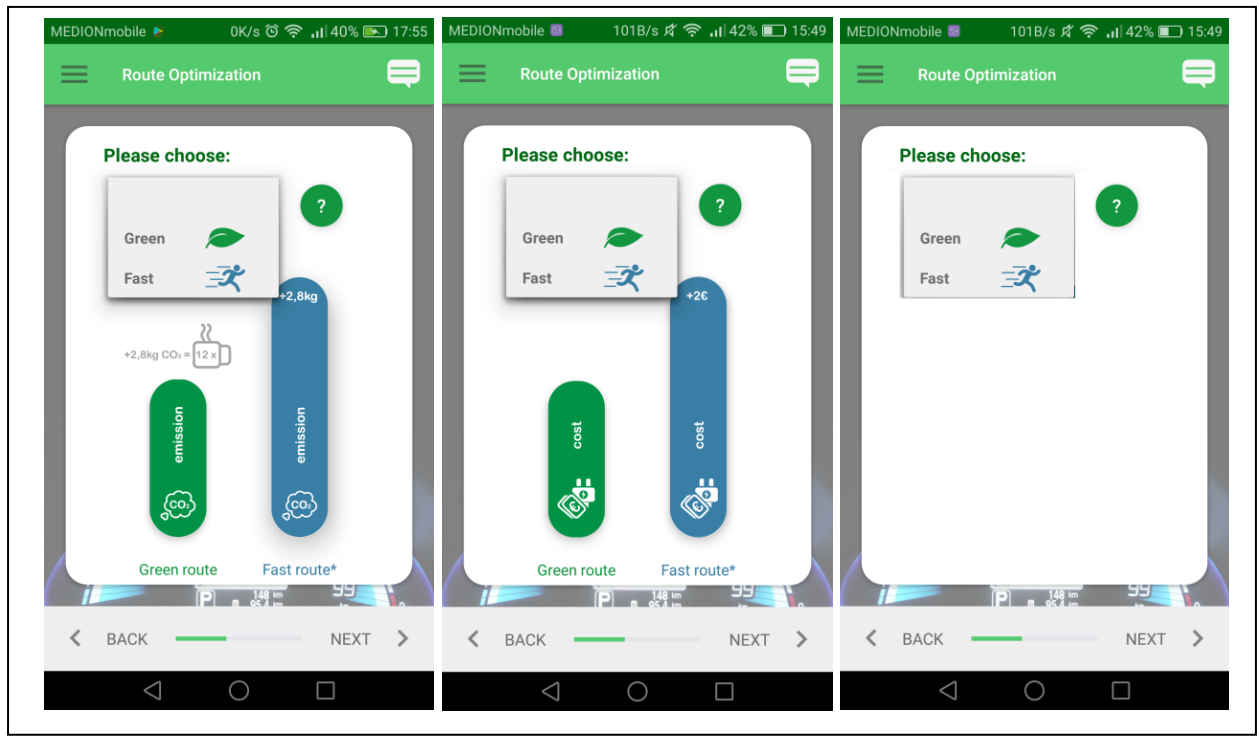

Figure 2. Electrific application, incentive screens. Left (2a): symbolic incentive screen displaying the additional emission cost for fast route choice. Middle (2b): financial incentive displaying additional financial cost of choosing the fast route. Right (2c): control condition without incentive.

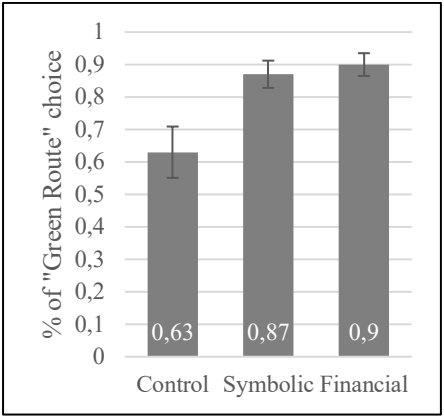

Figure 3. Means and SEs of green route choices made by participants in the three incentive conditions.

## Appendix – Scale Items

**EV items (5-point Likert scale, strongly disagree – indifferent – strongly agree)**

**Purchase intentions**

Overall, I think that my next private car will be fully electric.

**EV attractiveness**

Overall, I think EVs are attractive.

Overall, I think that EVs are the future.

**New Ecological Paradigm (5-point Likert scale, strongly disagree – indifferent – strongly agree)** (Stern et al., 1995)

When humans interfere with nature it often produces disastrous consequences.

The Earth is like a spaceship with very limited room and resources.

If things continue on their present course, we will soon experience a major ecological catastrophe.

When I drive a car, I contribute to pollution.

**Attitude functions (5-point Likert scale, strongly disagree – indifferent – strongly agree)**

I drive/would drive a fully electric vehicle, because

**Value attitude functions**

...it lets me express my values.

... I can be part of the sustainability movement

... it is healthier due to lack of fumes and pollution

... I can be environmentally friendly

**Utility attitude functions**

... it is easier to drive due to its quick acceleration and its silent motor

... it can save me time due to home/destination charging

… I have more control

... it is more fun to drive due to its quick acceleration and its silent motor

## Full model output

**H1 control vs treatments  &  H2 monetary vs symbolic**

Coefficients:

| | Estimate | Std. Error | z value | Pr(>\|z\|) | |
|---|---|---|---|---|---|
| (Intercept) | 0.60672 | 0.67520 | 0.899 | 0.36888 | |
| **h1_control_vs_treatments** | **1.18012** | **0.54568** | **2.163** | **0.03057** | * |
| h2_monetary_vs_symbolic | -0.98479 | 0.68875 | -1.430 | 0.15277 | |
| **gender** | **2.09110** | **0.77766** | **2.689** | **0.00717** | ** |
| NEP_mean_cent | -0.51745 | 0.62573 | -0.827 | 0.40826 | |
| att_fun_instrumental_cent | 0.33882 | 0.45923 | 0.738 | 0.46064 | |
| att_fun_enviro_cent | 0.30584 | 0.53804 | 0.568 | 0.56974 | |
| EVexp | -0.35117 | 0.60172 | -0.584 | 0.55949 | |
| h1_control_vs_treatments:gender | -0.49876 | 0.60223 | -0.828 | 0.40757 | |
| h1_control_vs_treatments:NEP_mean_cent | 0.04338 | 0.42235 | 0.103 | 0.91819 | |
| h1_control_vs_treatments:att_fun_instrumental_cent | -0.27229 | 0.30728 | -0.886 | 0.37555 | |
| h1_control_vs_treatments:att_fun_enviro_cent | 0.17171 | 0.34734 | 0.494 | 0.62106 | |
| h1_control_vs_treatments:EVexp | 0.08084 | 0.40703 | 0.199 | 0.84257 | |
| h2_monetary_vs_symbolic:gender | 1.35849 | 0.85218 | 1.594 | 0.11091 | |
| h2_monetary_vs_symbolic:NEP_mean_cent | -0.54071 | 0.79966 | -0.676 | 0.49892 | |
| h2_monetary_vs_symbolic:att_fun_instrumental_cent | -1.11845 | 0.59111 | -1.892 | 0.05847 | . |
| h2_monetary_vs_symbolic:att_fun_enviro_cent | 0.79791 | 0.71171 | 1.121 | 0.26224 | |
| h2_monetary_vs_symbolic:EVexp | -0.57955 | 0.76758 | -0.755 | 0.45022 | |

**Exporatory: magnitude**

Coefficients:

| | Estimate | Std. Error | z value | Pr(>\|z\|) |
|---|---|---|---|---|
| (Intercept) | 19.00221 | 3535.58171 | 0.005 | 0.996 |
| incentive_magnitude | 17.18074 | 3535.58197 | 0.005 | 0.996 |
| incentive_groupsymbolic | -17.53227 | 3535.58184 | -0.005 | 0.996 |
| gender | -16.46723 | 3535.58177 | -0.005 | 0.996 |
| NEP_mean_cent | 0.12325 | 1.02413 | 0.120 | 0.904 |
| EVexp | 15.67584 | 2367.71915 | 0.007 | 0.995 |
| att_fun_instrumental_cent | -0.03104 | 0.52640 | -0.059 | 0.953 |
| att_fun_enviro_cent | 0.26577 | 0.61374 | 0.433 | 0.665 |
| incentive_magnitude:incentive_groupsymbolic | -18.70638 | 3535.58212 | -0.005 | 0.996 |
| incentive_magnitude:gender | -17.20272 | 3535.58198 | -0.005 | 0.996 |
| incentive_magnitude:NEP_mean_cent | -0.31954 | 1.17031 | -0.273 | 0.785 |
| incentive_magnitude:EVexp | 15.60867 | 2367.71946 | 0.007 | 0.995 |
| incentive_groupsymbolic:gender | 19.11231 | 3535.58199 | 0.005 | 0.996 |
| incentive_groupsymbolic:NEP_mean_cent | -1.00382 | 1.76651 | -0.568 | 0.570 |
| incentive_groupsymbolic:EVexp | -17.22660 | 2367.71945 | -0.007 | 0.994 |
| incentive_magnitude:incentive_groupsymbolic:gender | 17.19117 | 3535.58224 | 0.005 | 0.996 |
| incentive_magnitude:incentive_groupsymbolic:NEP_mean_cent | 0.41336 | 2.02064 | 0.205 | 0.838 |
| incentive_magnitude:incentive_groupsymbolic:EVexp | -13.48437 | 2367.71993 | -0.006 | 0.995 |